\documentclass[preprint,showpacs,floats,letterpaper,floatfix,,groupedaddress,eqsecnum]{revtex4}
\usepackage{subfigure}

\usepackage{amssymb,amsmath}
\usepackage[dvips]{graphicx}

\usepackage{dcolumn,epsfig}

\begin{document}

\title{Particle dynamics and Stochastic Resonance in Periodic potentials}
\author{Shantu Saikia$^1$}
\email{shantusaikia@anthonys.ac.in}
\affiliation{$^1$St. Anthony's College, Shillong-793001, Meghalaya, India}

\begin{abstract}
We have studied the dynamics of a particle in a periodically driven underdamped periodic 
potential. Recent studies have reported the occurrence of Stochastic Resonance (SR) in 
such systems in the high frequency regime, using input energy per period of external drive
as a quantifier. The particle trajectories in these systems can be in two dynamical states 
characterised by their definite energy and phase relation with the external drive. SR is
due to the noise assisted transition of the particles between these two states. We study 
the role of damping on the occurrence of SR. We show that a driven underdamped periodic system
exhibits SR only if the damping is below a particular limit. To explain this we study the
syatem in the deterministic regime. The existence of the two dynamical states in the 
deterministic regime is dependent on the amount of damping and the amplitude od external 
drive. We also study the input energy distributions and phase difference of the response 
amplitude with the external drive as afunction of the friction parameter.

\end{abstract}

\date{\today}
\maketitle

\section{Introduction}
	Stochastic Resonance (SR) is a phenomenon exhibited by non-linear systems in
 which the system shows an enhanced response to an external periodic forcing in the 
presence of an optimal amount of noise \cite{Ben, Gam, Well}. Initially thought of
 as an explanation for the recurring ice ages on earth \cite{Ben}, presently SR finds
lots of applications in diverse biological \cite{Moss, Maddox, Hanggi, Ghosh, Douglass, 
Collins, Gluckman, Simonotto, Shuhei}  and physical systems \cite{Fuave, Mantegna, Murali, 
Roy, Mohanty}. Extensive theoritical and experimental work has been done on SR \cite{Gam, 
Well}.
\par Initially considered to be a phenomenon exhibited only by bistable systems 
\cite{Gam, Moss}, recently there has been some studies of non-conventional SR in 
monostable and multistable systems \cite{Stocks, Dykman, Arathi, Wies, Kauf, Lin, 
Dan, Fron, Hu, Cass, Gitt, Bao, Marc, Kall, Kim,Reen2, Schia, Zhang, Saikia, Reen1, Liu}. 
In particular SR in periodic potential systems has been of interest to researchers 
both in the overdamped \cite{Lin, Dan, Fron, Hu, Cass, Gitt, Bao} and underdamped limit 
\cite{Marc, Kall, Kim,Reen2, Schia, Zhang, Saikia, Reen1, Liu}. 

\par Earlier studies have established the phenomenon of SR in periodic washboard 
potentials \cite{Dan, Fron, Hu, Cass, Gitt, Bao, Marc, Kall, Reen2}. However, the phenomenon
of SR in periodic potentials without any bias has been of recent interest \cite{Kim,Reen2,
Schia, Zhang, Saikia, Reen1, Liu}. Though there was some controversy as to whether
SR occurs in an underdamped periodic potential \cite{Kim}, recent studies have conclusively
proved the occurrence of SR in such systems \cite{Reen2, Schia, Zhang, Saikia, Reen1, Liu}.
Underdamped particle motion in periodic potentials has practical relevance in understanding
diverse physical phenomenon \cite{Risk} such as electrical conductivity of superionic 
conductors \cite{Fulde}, adatom motion on crystal surfaces \cite{Laca}, resistively 
coupled shunted junction (RCSJ) model of Josephson junctions \cite{Falco} etc.

\par In an experimental work \cite{Schia}, the SR phenomenon in an underdamped periodic 
potential system was studied by taking a dissipative optical lattice as
the spatially periodic system. They used laser fields to create the periodic potential
and produce the stochastic process of optical pumping. The friction experienced by the
atoms was kept low so that they exhibited underdamped motion. Using this model, Schiavoni
et. al. observed the phenomenon of SR on the Brillouin propagation modes of the dissipative
optical lattice. They also gave a theoretical account of their experimental findings.
Zhang et.al \cite{Zhang} studied the mobility and diffusion of underdamped Brownian
particle in a two-dimensional periodic potential. They observed that the effect of two
dimensions leads to the observance of SR in the system. The output quantities like 
signal amplification and diffusion rate shows a peaking behaviour with temperature
which otherwise is absent in the one dimensional potential.

\par In an earlier work, using the average input energy per period of drive and hysteresis
 loop area as quantifiers of SR, it was shown that the phenomenon of SR occurs in a
periodically driven underdamped periodic potential in the high-frequency regime of the
external forcing \cite{Saikia}.
The phenomenon of SR was observed with a periodic potential system which was symmetric 
and also when an asymmetry was introduced into the system in the form of a space dependent 
friction coefficient. Such a spatially inhomogeneous periodic potentials system have
been studied earlier to obtain particle current in the overdamped \cite{Pareek, DanN} as 
well as in the underdamped \cite{SaikiaN} regime.
In this work, it is shown that the particles in the system can be in two distinct
dynamical states of trajectories with distinct amplitude and phase relationship with the
external drive. At low temperatures, the two dynamical states are stable with fixed 
energies. However with the rise of temperature, transition occurs between these two states.
The phenomenon of SR was attributed to the noise assisted transitions 
that the particles make between these two states, synchronised with the external drive. 
In a subsequent work \cite{Reen1}, the role of the driving amplitude on the occurrence of 
SR was studied. It was shown that there is a particular range of driving amplitude for which 
the system exhibits the phenomenon of SR. This is because the relative stability of the
two dynamical states of the particle depends on the amplitude of the external drive.
Similar SR phenomenon was also observed in a bistable periodic potential and washboard
potential \cite{Reen2}. All the above works on SR \cite{Reen2, Saikia, Reen1} were with 
Gaussian white noise. In a recent work \cite{Liu}, the effect of colored noise on the 
phenomenon of SR in the underdamped motion of a particle in a periodic potential was
addressed. Colored noise was shown to affect the transitions between the two dynamical
states and thus SR. Also it was observed that, as the correlation time increases,
the SR peak shifts and the sharpness of the peak decreases.

\par In this work we study the role of damping on SR in a periodically driven underdamped 
periodic  potential with Gaussian white noise, using the same model as in Ref.  
\cite{Saikia, Reen1}. We use input energy per period of external drive $F(t)$ as a quantifier 
of SR. The input energy per period of external drive or the work done on the system per 
period of drive is calculated using the stochastic energetics formulation of Sekimoto 
\cite{Seki}. This quantity has been found to be a good quantifier for SR in earlier 
studies for bistable systems \cite{Iwai, Dan1, Saikia1} as well as periodic potentials 
\cite{Reen2, Saikia, Reen1, Liu}. 
We show that in a driven underdamped periodic system, the amount of damping present in the
medium plays a crucial role on the occurrence of SR. In such a system the particle 
trajectory, $x(t)=x_0cos(\omega t +\phi)$, can exist in two definite dynamical states,
with a distinct phase difference $\phi$ with the external drive $F(t)$; the low amplitude 
$\it{in-phase}$ state, with $\phi=\phi_1$ which is almost in phase with the external drive, 
and the higher amplitude, $\it{out-of-phase}$ state, with $\phi=\phi_2$, having higher 
phase difference with the external drive. The $\it{in-phase}$ state has lower energy 
associated with it and the $\it{out-of-phase}$ state has higher energy. The occurrence of 
SR is due to the noise assisted transitions that the particles make between these two states 
\cite{Reen2, Saikia, Reen1, Liu}. To explain the observed dependence of SR on damping we 
study the stability of the two dynamical states of trajectories with respect to the
different parameters like temperature, amplitude of drive and the friction coefficient. 
We study the input energy distributions at different temperatures across the SR peak for
various values of the friction coefficient, to chararcterise SR as has been done in earlier
studies \cite{Reen2,Saikia, Saikia1}. From the average response of the particle to the 
external drive ${\bar x(t)}={\bar x_0}cos(\omega t +{\bar \phi})$, we calculate the
average phase difference $\bar{\phi}$ from the $F(t)-\bar{x(t)}$ hysteresis loops.  
 We study the variation $\bar{\phi}$ with different values of damping. The role of 
the average phase difference $\bar{\phi}$ of the response amplitude 
$\bar {x(t)}$ and the external forcing $F(t)$ on SR has been explored  earlier 
\cite{Saikia, Iwai, Gamma1, Dykman1}.

\par In section II we discuss the model used. Section III contains the detailed discussion of
our results while in Section IV we conclude our results.

\section{The model}

In this work, we consider the underdamped motion of a particle in a 
periodic potential $V(x)=-V_0 \sin(kx)$ which is symmetric in space 
and having a period of $2\pi$.  The system is driven 
periodically by an external forcing $F(t)$=$F_0\cos(\omega t)$. The constants
$V_0$ and $F_0$ are the amplitudes of the potential and the driving force respectively.

\par The dynamics of a particle of mass $m$ moving in a periodic potential
 $V(x)=-V_0\sin(kx)$ in a medium with friction coefficient $\gamma $ driven by an
extenal periodic forcing $F(t)$, in the presence of random fluctuations is described by the 
Langevin equation,

\begin{equation}
m\frac{d^{2}x}{dt^{2}}=-\gamma\frac{dx}{dt}-\frac{\partial{V(x)}}{\partial
x}+F(t)+\sqrt{\gamma T}\xi(t).
\end{equation} 
In the above equation, the temperature $T$ is in units of the Boltzmann constant $k_B$. 
$\xi (t)$ represents the random fluctuations in the system satisfying the statistics:
 $<\xi (t)>=0$, and 
$<\xi(t)\xi(t^{'})>=2\delta(t-t^{'})$. 
The above equation can be written in dimensionless units by setting $m=1$, $V_0=1$, 
$k=1$. Using the same symbols for the reduced variables, the dimensionless form
of the Langevin equation becomes

\begin{equation}
\frac{d^{2}x}{dt^{2}}=-\gamma\frac{dx}{dt}
+cos x +F(t)+\sqrt{\gamma T}\xi(t).
\end{equation}

In the above equation too, $\xi(t)$ obeys the same statistics as before.
 For studying the particle dynamics in the deterministic regime, $T$ is set 
equal to $0$.

\section{Numerical Results}
	The underdamped Langevin's equation (Eq. 2) is solved numerically using 
$2^{nd}$ order Heun's method \cite{Nume}, by treating it as an initial value problem.
We take an integration-time step $\Delta t=0.001$. In the deterministic regime 
\cite{Saikia2} and also in the low temperature, low amplitude regime \cite{Saikia}, 
the dynamics of the particle is sensitively dependent on the initial conditions, 
$x(0)=x(t=0)$ and $v(0)=v(t=0)$. So for physically relevant results, ensemble 
averaging is done over an ensemble of 100 particles starting with different initial 
positions, $x_i(0)$, $i=1, 2, ...... 100$, taken from a uniformly spaced grid between 
the two consecutive peaks of the periodic potential and initial velocity $v=0$ 
\cite{Mateos}. 

\par The input energy, or work done on the system per period of the external drive, $E_i$,
 is calculated using the stochastic energetics formulation of Sekimoto \cite{Seki} as
\begin{equation}
E_i(t_0,t_0+\tau)=\int_{t_0}^{t_0+\tau}\frac{\partial U(x(t),t)}{\partial t}dt,
\end{equation}
where, $\tau$ is the period of the external drive, the potential $U(x(t),t)=V(x)-xF(t)$
, and $V(x)=-\sin(x)$, $F(t)=F_0\cos(\omega t)$. For our calculations we consider
$F_0=0.2$ and $\tau=8.0$ unless otherwise stated. The value of $\tau=8.0$ is in the 
high frequency regime close to the natural frequency at the bottom of the potential
wells.

\par To find the average input energy per period of external drive 
$\langle \overline{E_i}\rangle$,
averaging is first done over all the periods of one trajectory as

\begin{equation}
\overline{E_i}= \frac{\sum_{n=N_t}^{N_f}{E_i}_n}{N_f-N_t}
\end{equation}
The number of periods, $N_f$, taken in a trajectory ranges between
$10^5$ to $10^7$, as required. $N_t$ is the initial number of transients
which are removed. $\langle\overline{E_i}\rangle$ is then calculated by averaging 
$\overline{E_i}$ over all the trajectories.

\par The results of our numerical calculations are presented in the following
sections.

\subsection{Input energy versus temperature for different damping}

\begin{figure}[h]
\begin{center}
\epsfig{file=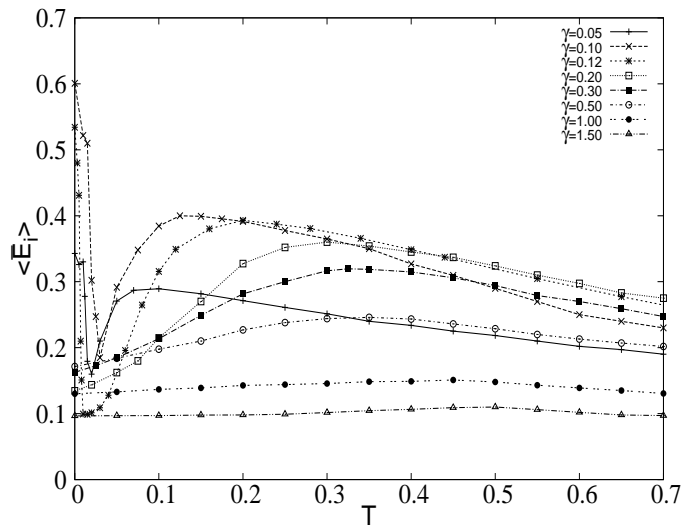, angle=-90, width=9.0cm,totalheight=7cm}
\caption{Plot of $\langle \overline E_i \rangle$  versus $T$, for different values of the 
friction coefficient $\gamma$; $F_0 = 0.2$ and $\tau =8$.}
\end{center}
\end{figure}

\par Fig.1. shows the main result of our work. The input energy per 
period or the work done per period over external drive $\langle \overline{E_i}\rangle$ 
is plotted versus the temperature $T$ for different values of the friction coefficient 
$\gamma$ in 
the medium. It is observed that for lower values of $\gamma$, the 
$\langle \overline{E_i}\rangle$ versus
$T$ curve shows a peaking behaviour which is a signature of SR 
\cite{Reen2, Saikia, Reen1, Liu}. However as $\gamma$ increases, the phenomenon
of SR becomes negligible and is almost absent for $\gamma=1.50$.
Also the sharpness of resonance decreases with the increase in the value of $\gamma$.

\begin{figure}[h]
  \centering
  \subfigure[]{\includegraphics[width=5.5cm,height=6.7cm,angle=-90]{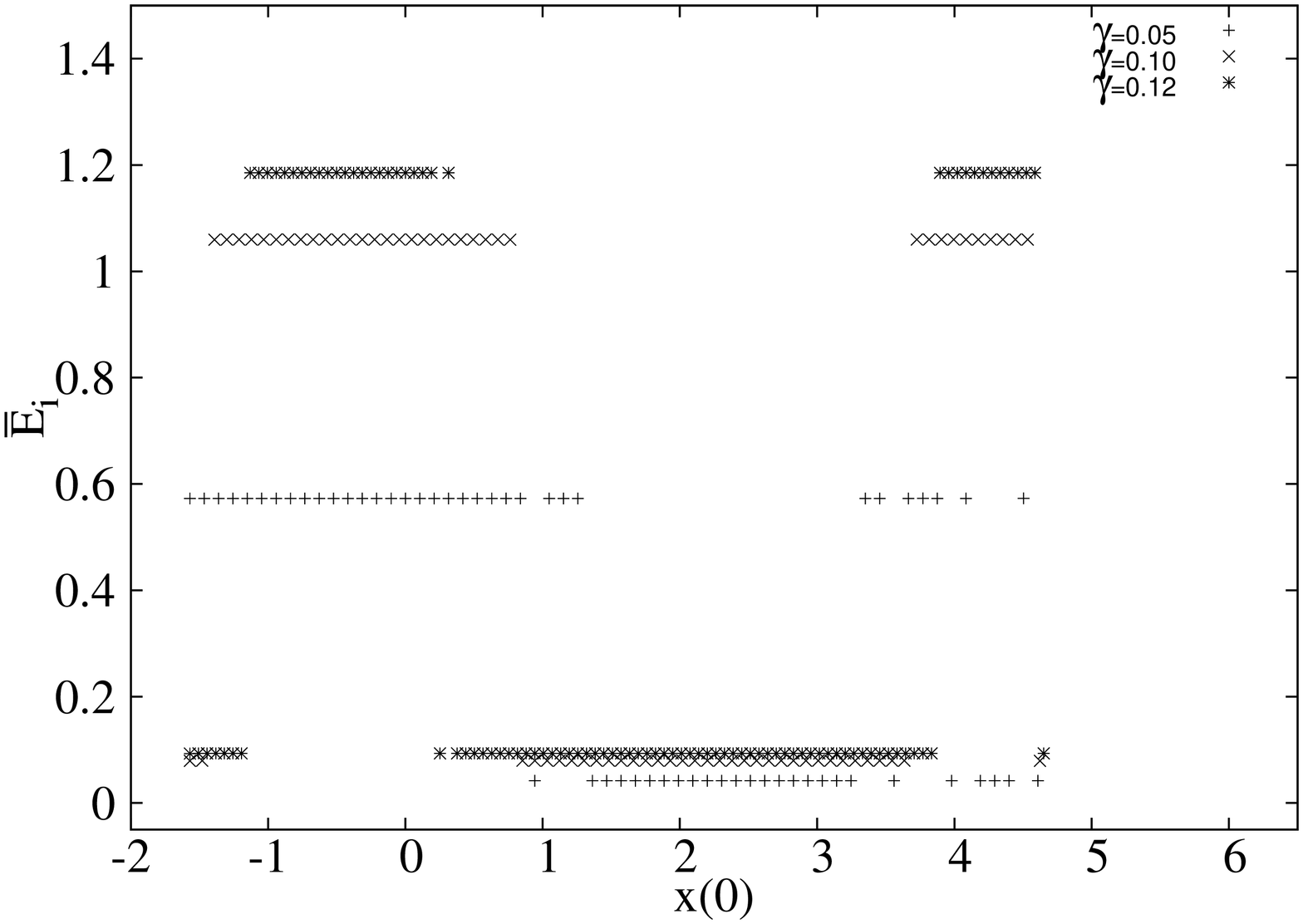}}
\hspace{0.01cm}
\subfigure[]{\label{fig:edge-c}\includegraphics[width=5.5cm,height=6.7cm,angle=-90]{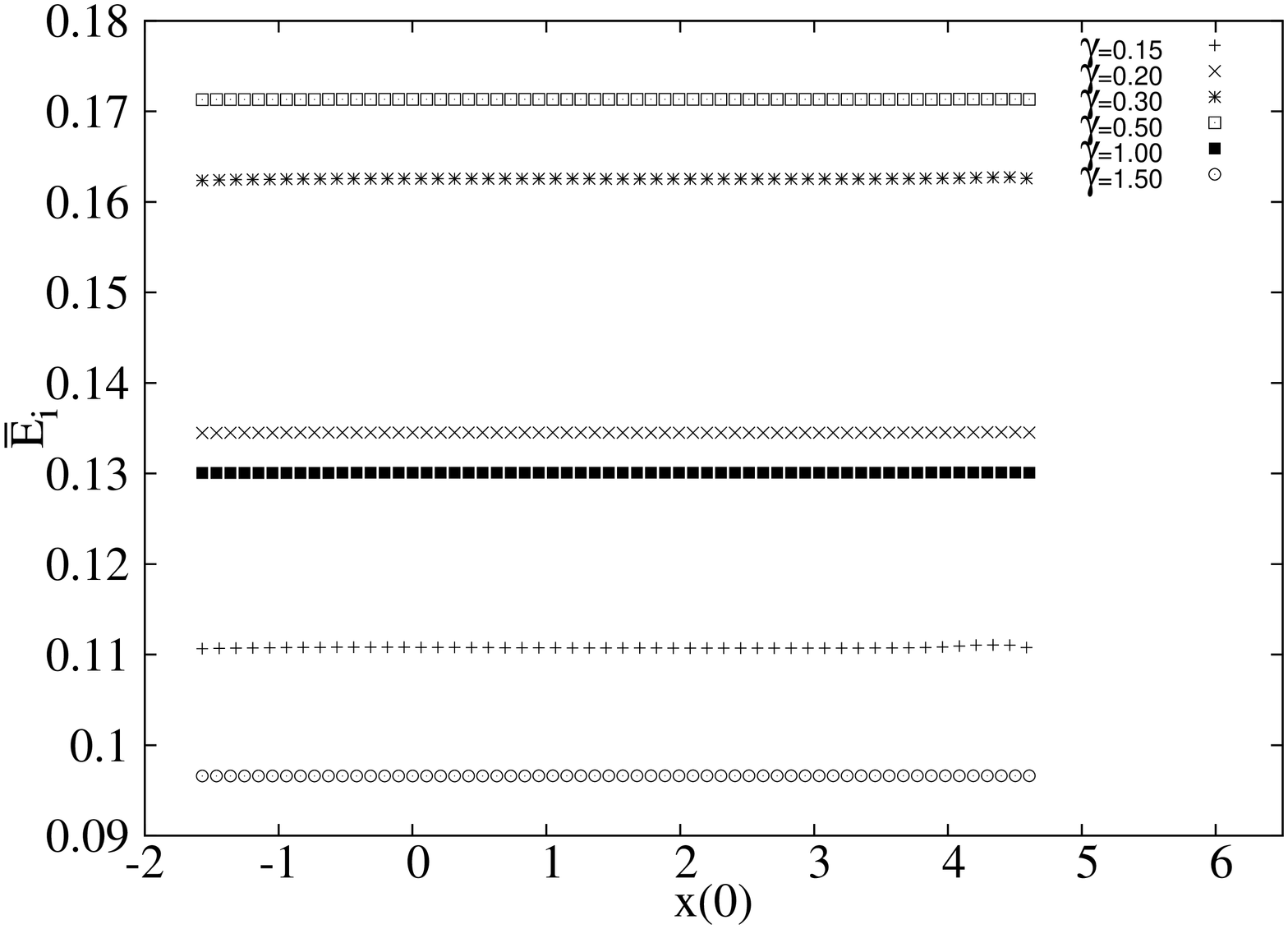}}
\caption{Plot of $\overline{E_i}$ versus $x(0)$ for different values of the friction
coefficient $\gamma$; $F_0 = 0.2$ and $\tau = 8.0$.}
\label{fig:edge}
\end{figure}

\par There is a significant change in the nature of the $\langle\overline{E_i}\rangle$ 
versus $T$ curves at lower temperatures as $\gamma$ changes. For $\gamma = 0.05, 0.10$ and 
$0.12$, as temperature increases, $\langle\overline{E_i}\rangle$  
attains a minimum and then increases to attain a peak at an intermediate $T$. However for 
$\gamma = 0.2, 0.3, 0.5$ and $1.0$ the minima observed at lower $T$ is missing.

 \begin{figure}[h]
  \centering
  \subfigure[]{\includegraphics[width=5.5cm,height=6.7cm,angle=-90]{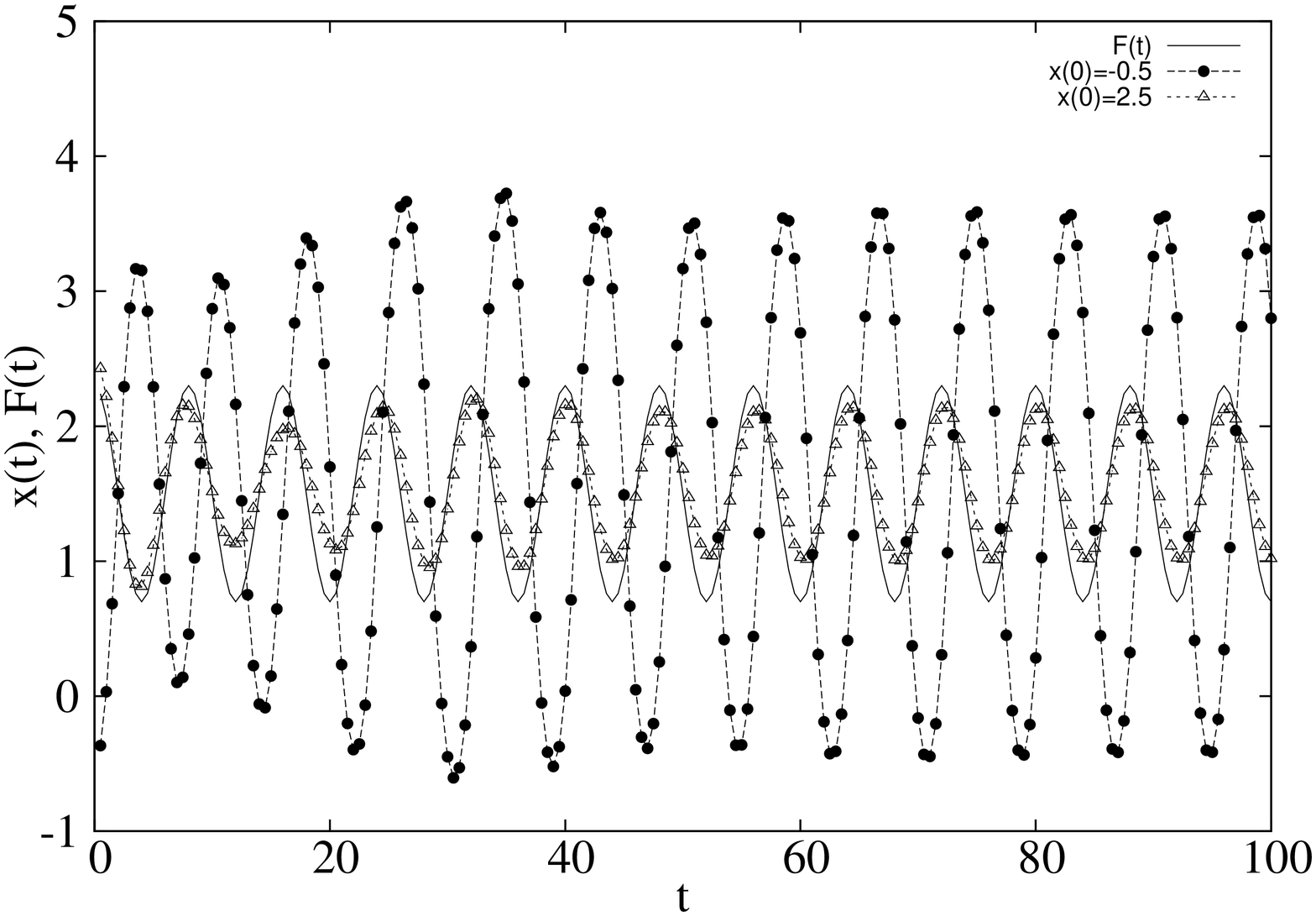}}
\hspace{0.01cm}
\subfigure[]{\label{fig:edge-c}\includegraphics[width=5.5cm,height=6.7cm,angle=-90]{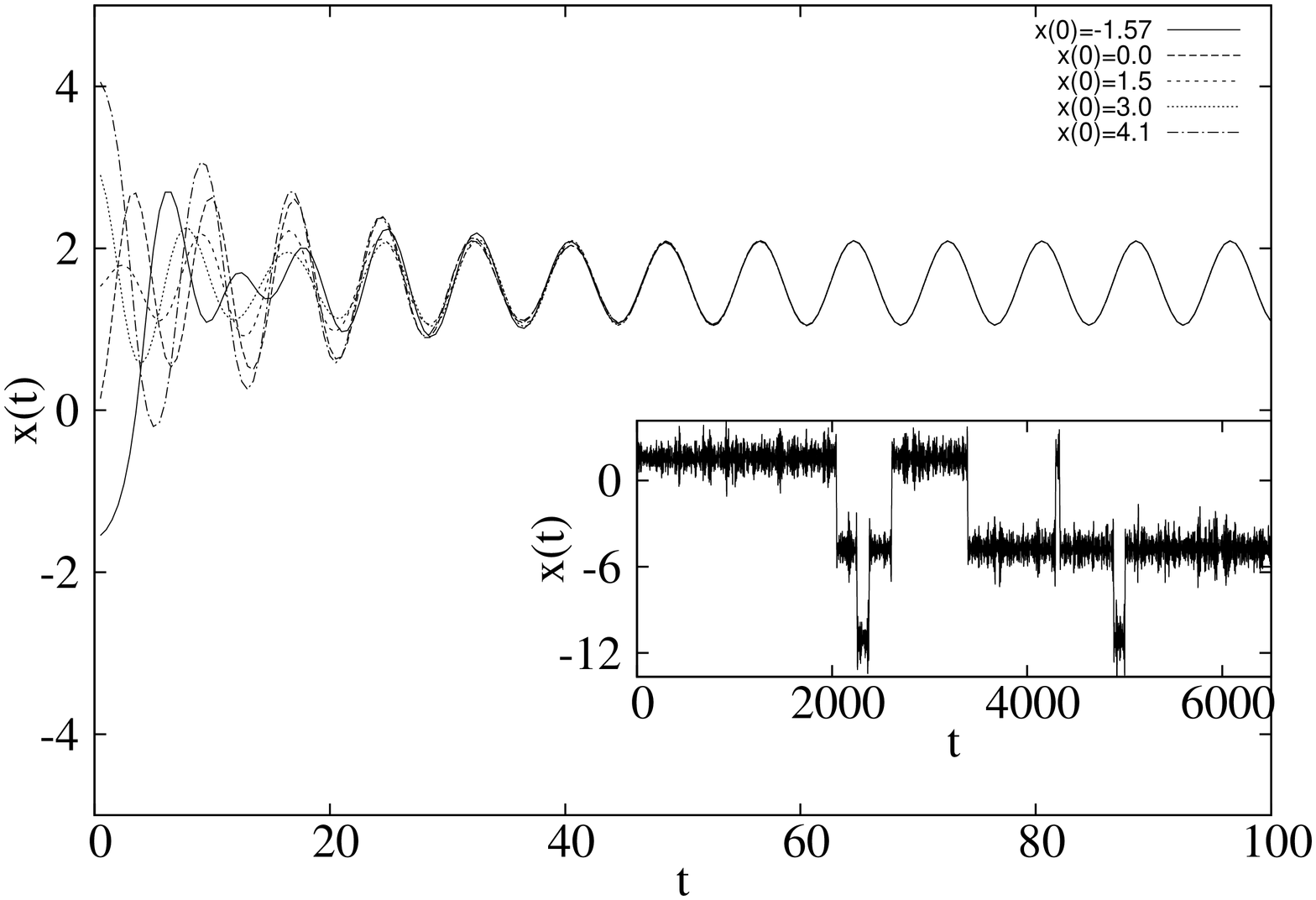}}
\caption{Plot of $x(t)$ and $F(t)$ for different $x(0)$ with $T=0$ for $\gamma = 0.1$(a)
and for $\gamma=0.2$ (b). Inset of (b) shows plot of $x(t)$ for $\gamma=0.2$ with $T=0.2$;
$x(0)=-1.57$, $F_0=0.2$, $\tau = 8.0$.}
\label{fig:edge}
\end{figure}

\subsection{Particle dynamics in different parameter regimes}

\par To explain this observed behaviour, we study the particle dynamics in the deterministic
regime. Fig. 2 shows the plot of average input energy per period in a trajectory, 
$\overline{E_i}$ versus the initial position $x(0)$ for different values of $\gamma$.
For lesser damping, in the deterministic limit (and also for lower
temperatures \cite{Saikia}), the particles can exist in two distinct dynamical states
of trajectories depending on the initial positions $x(0)$ (Fig. 2a). These dynamical states are 
distinctly characterised by the phase difference $\phi$ between the particle trajectory
$x(t)=x_0cos(\omega t +\phi)$ and the external periodic drive $F(t)=F_0cos(\omega t)$.
 The $\it{in-phase}$ has lower amplitude and hence lesser energy and while the 
$\it{out-of-phase}$ state has higher amplitude and hence higher energy. 

\begin{figure}[h]
\begin{center}
\epsfig{file=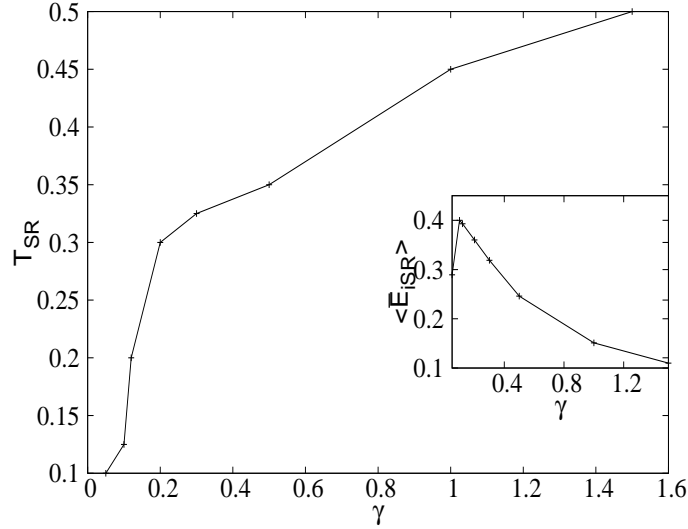, angle=-90, width=9.0cm,totalheight=7cm}
\caption{Plot of $T_{SR}$  versus $\gamma$. Inset shows
the variation of $\langle \overline{E_{iSR}} \rangle$ with $\gamma$; $F_0 = 0.2$ and $\tau =8$}
\end{center}
\end{figure}

Fig. 3a shows the in-phase 
state of trajectory with $x(0)=2.5$ and the high amplitude out-of-phase trajectory with 
$x(0)=0.5$ in the deterministic limit ($T=0$). The presence of the high energy state 
leads to higher average input energy at $T=0$ or at low $T$. As $T$ increases, all the 
particles come down to the low energy state \cite{Saikia}. Thereafter, with the rise of 
temperature the particles make transitions between the two states. The average input energy 
peaks at a value of $T$, where optimum transitions occur, synchronised with the external drive.
 For higher damping, all the particles for a particular $\gamma$ value, are in the same 
energy state at $T=0$ (and also at lower $T$), irrespective of the initial conditions (Fig. 2b).
$\overline{E_i}$ comparatively lower than the high energy states of the low damping regime. 
A look into the particle trajectories reveals that when damping is high, the particles
looses sensitivity to the initial conditions and all the particles after sometime goes to the
same state of motion with lesser amplitude and hence energy (main Fig. 3b). 
This leads to low $\langle \overline{E_i}\rangle$ at $T=0$ or low $T$.  
Though at low $T$, the two dynamical states are not there for higher damping, as
$T$ increases, the particle makes transitions between a low amplitude and high amplitude 
motion (Inset of Fig. 3b). The peaking of $\langle\overline{E_i}\rangle$ 
at an optimum $T$ is therefore apparently due to the synchronised transitions between the
two types of particle motion.
\begin{figure}[h]
  \centering
  \subfigure[]{\includegraphics[width=6cm,height=6.7cm,angle=-90]{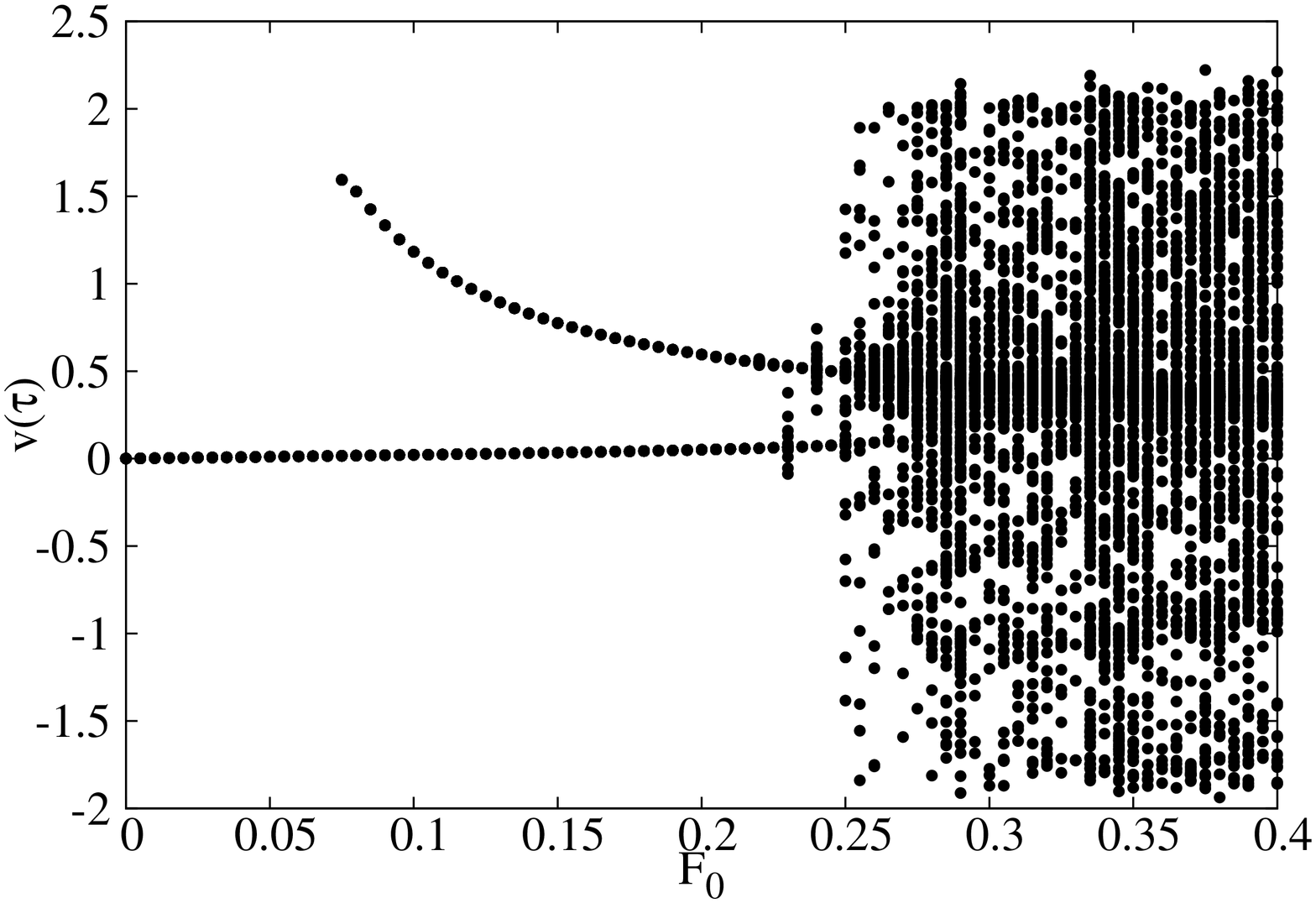}}
\hspace{0.01cm}
\subfigure[]{\includegraphics[width=6cm,height=6.7cm,angle=-90]{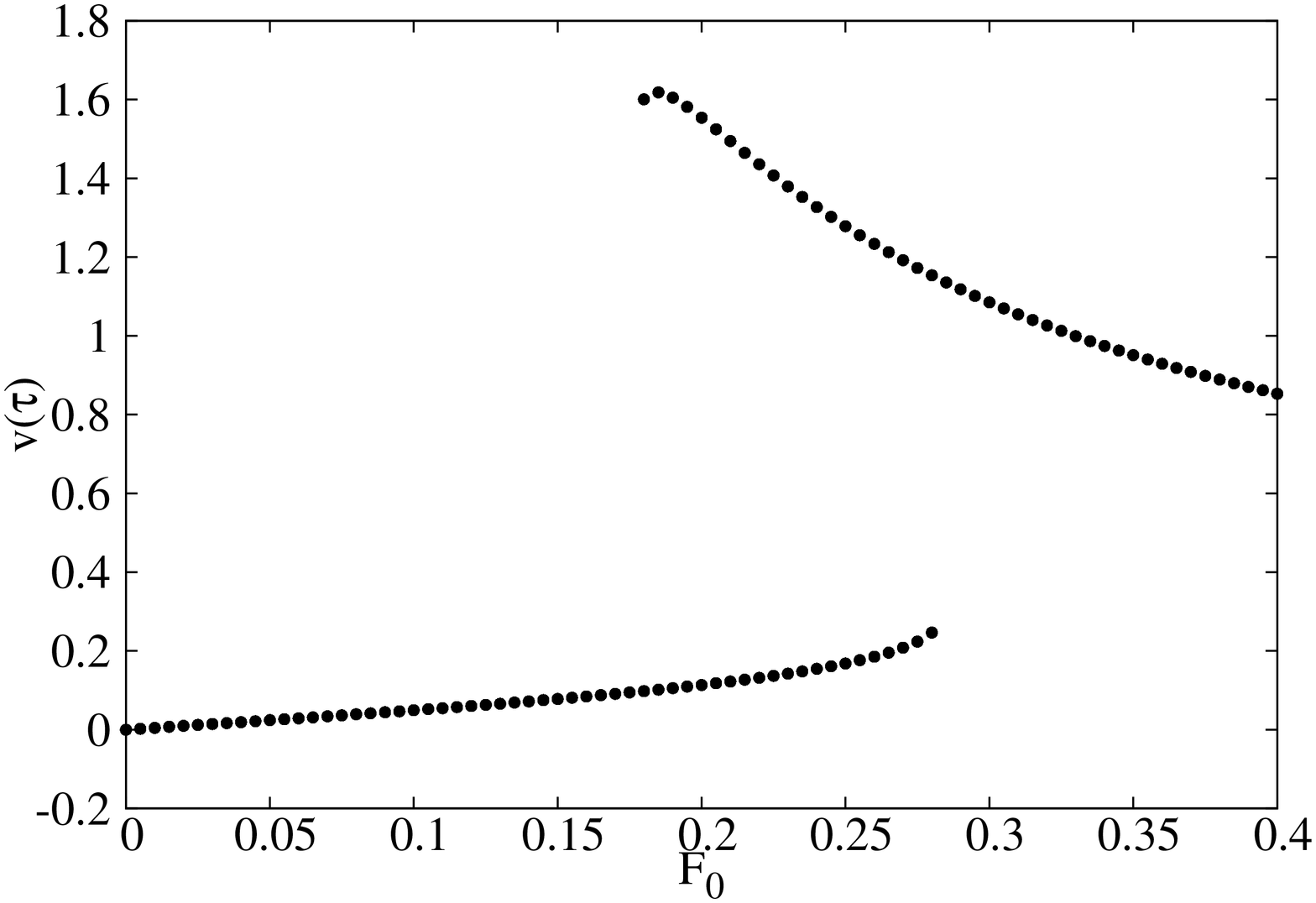}}
\hspace{0.01cm}
\subfigure[]{\label{fig:edge-c}\includegraphics[width=6cm,height=6.7cm,angle=-90]{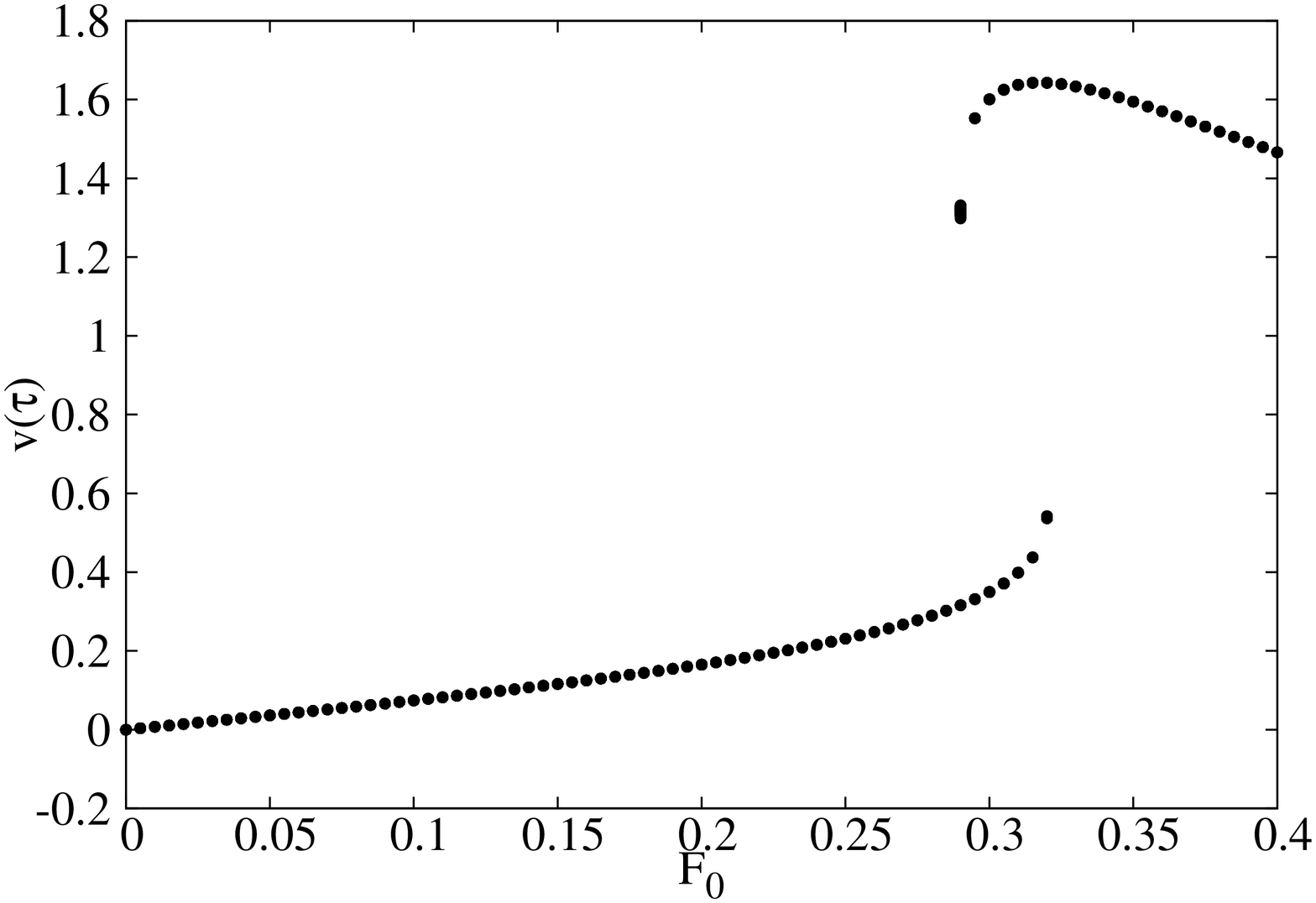}}
  \subfigure[]{\includegraphics[width=6cm,height=6.8cm,angle=-90]{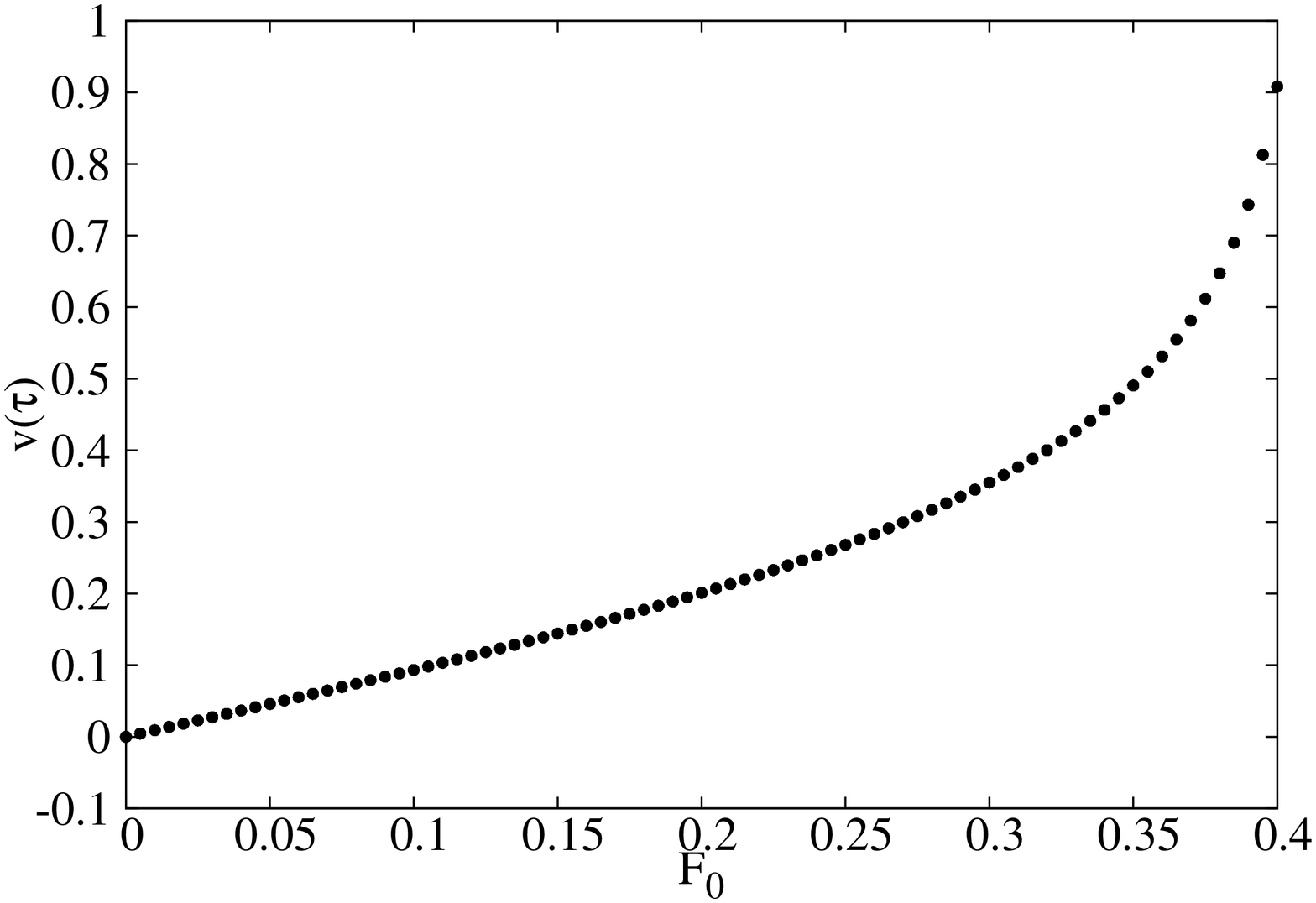}}
\hspace{0.01cm}
\caption{Bifurcation diagrams for different values of $\gamma$ in the deterministic
limit ($T=0$); $\gamma=0.05$ (a), $\gamma=0.12$ (b), $\gamma=0.2$ (c) and $\gamma=0.3$ (d);
 $\tau = 8.0$.}
\label{fig:edge}
\end{figure}

The value of temperature $T=T_{SR}$ at which the average input energy 
$\langle \overline{E_i} \rangle$ peaks, increases as the damping increases (Fig. 4). 
Significantly, the value of average input energy at the peak of the SR curve,
is maximum for an optimum value of damping (Inset of Fig. 4). 

\par The observed shift in the SR peak to higher $T$ for higher $\gamma$ is due to the 
fact that when damping is high, the optimum synchronisation 
of transistion between the two dynamical states of the particle occurs at higher
values of $T$.
\par At still higher values of $\gamma$, the particle motion reaches the overdamped regime,
 leading to the absence of the peaking behaviour.

\subsection{Bifurcation diagrams}

\par Fig. 5 shows the bifurcation diagram in the deterministic regime for different values 
of damping. These are 
obtained by recording the particle velocity $v(T_p)$ at times $T_p = \tau$
equal to the period of the external drive and plotting them as a function of a control
parameter, here $F_0$, the amplitude of the external drive \cite{Saikia2}. For the value 
of amplitude $F = 0.2$, considered for Fig. 1, the bifurcation diagram clearly shows 
the existence of two regular trajectories - one with higher velocity and the other with 
lesser velocity for $\gamma = 0.05, 0.10, 0.12$. 
\begin{figure}[h]
  \centering
  \subfigure[]{\includegraphics[width=6cm,height=6.8cm,angle=-90]{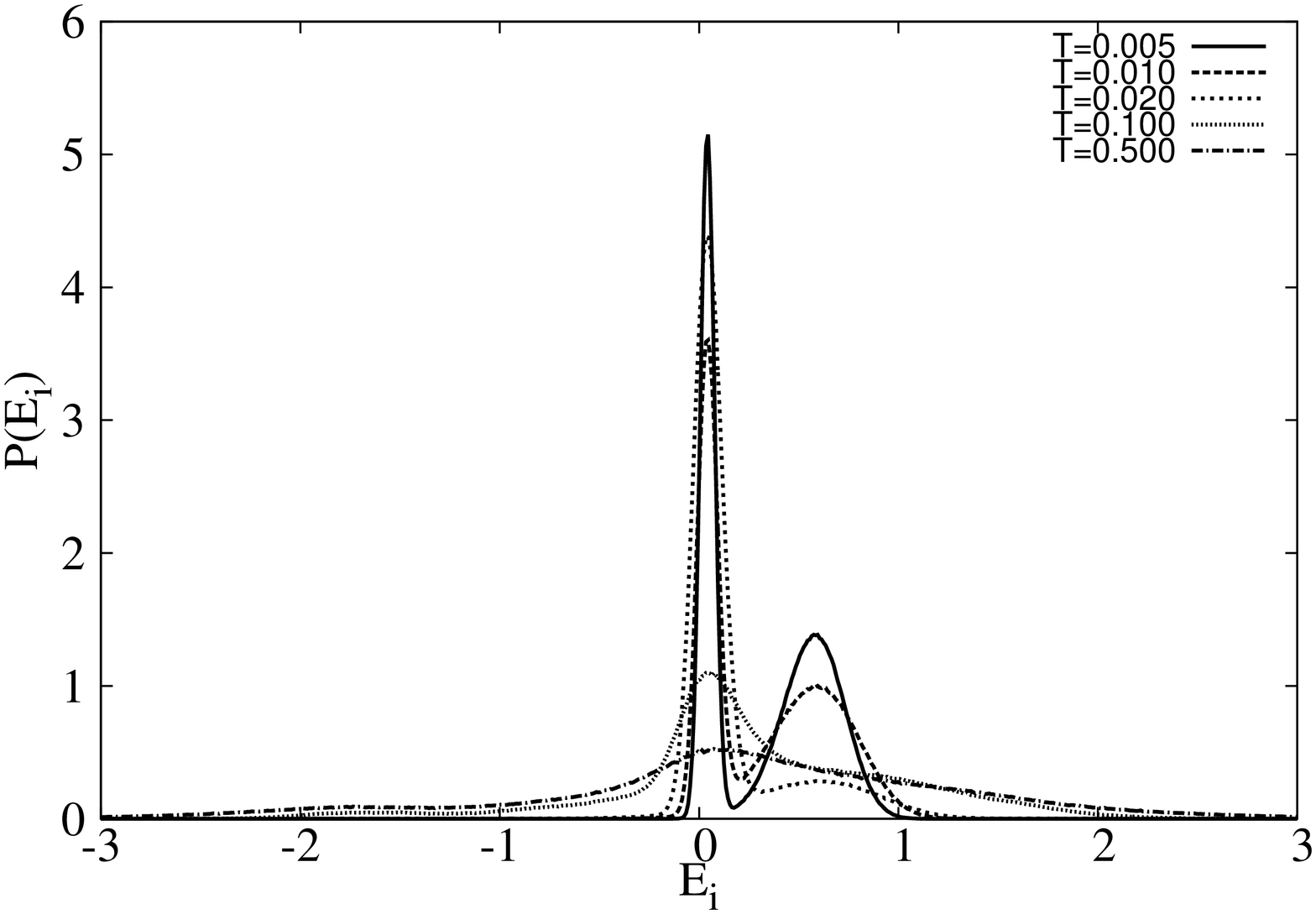}}
\hspace{0.01cm}
\subfigure[]{\includegraphics[width=6cm,height=6.8cm,angle=-90]{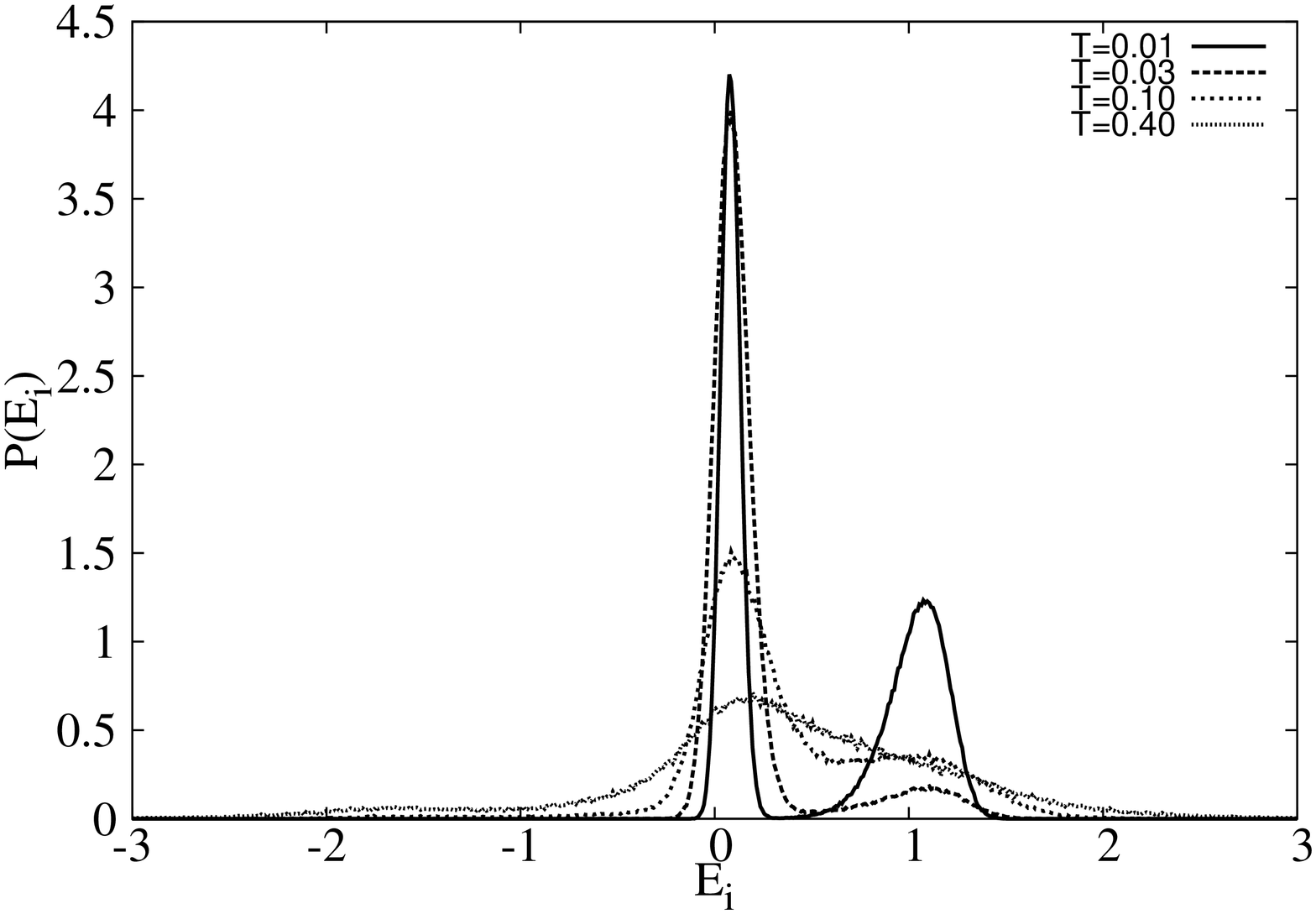}}
\hspace{0.01cm}
\subfigure[]{\label{fig:edge-c}\includegraphics[width=6cm,height=6.8cm,angle=-90]{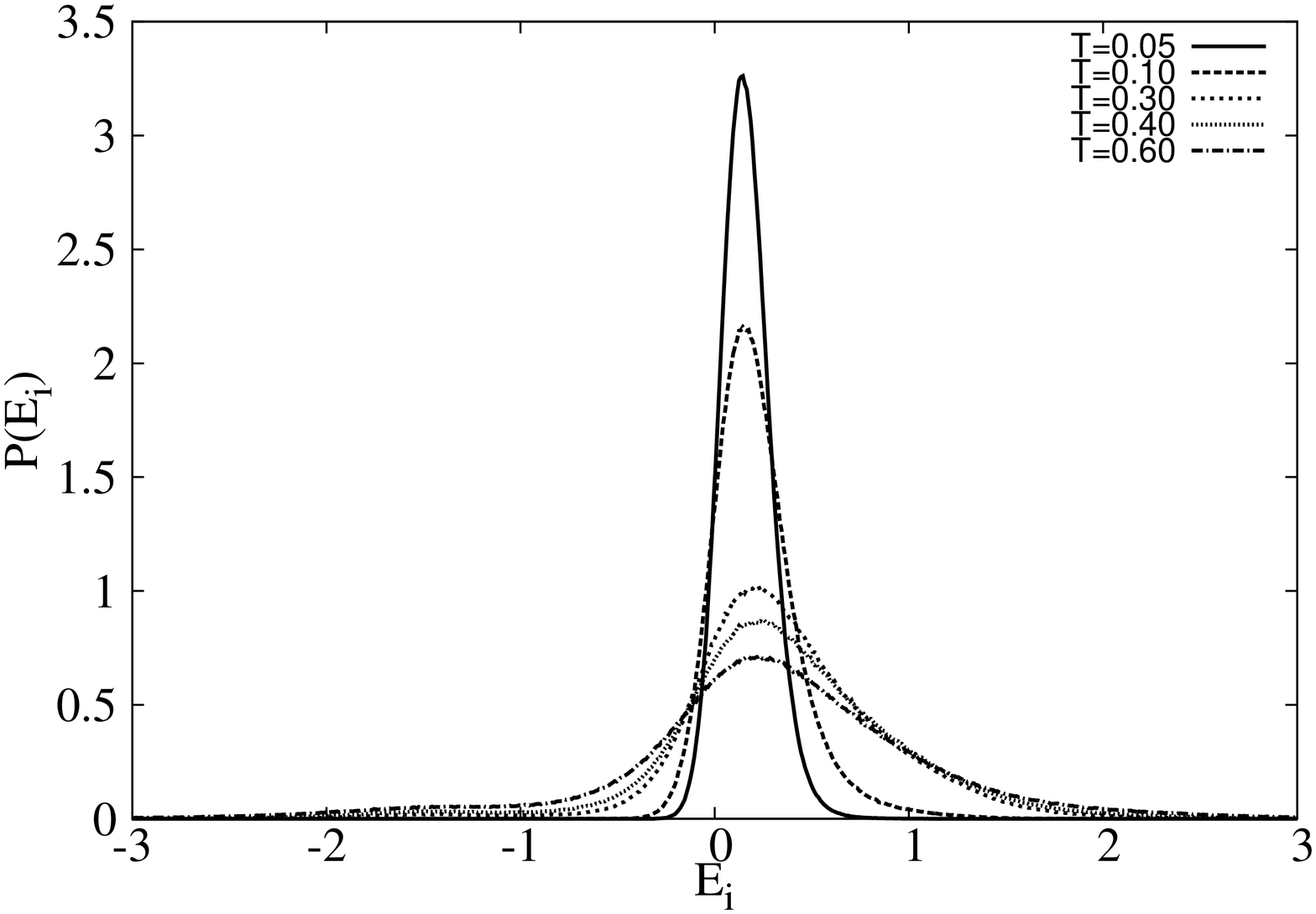}}
  \subfigure[]{\includegraphics[width=6cm,height=6.9cm,angle=-90]{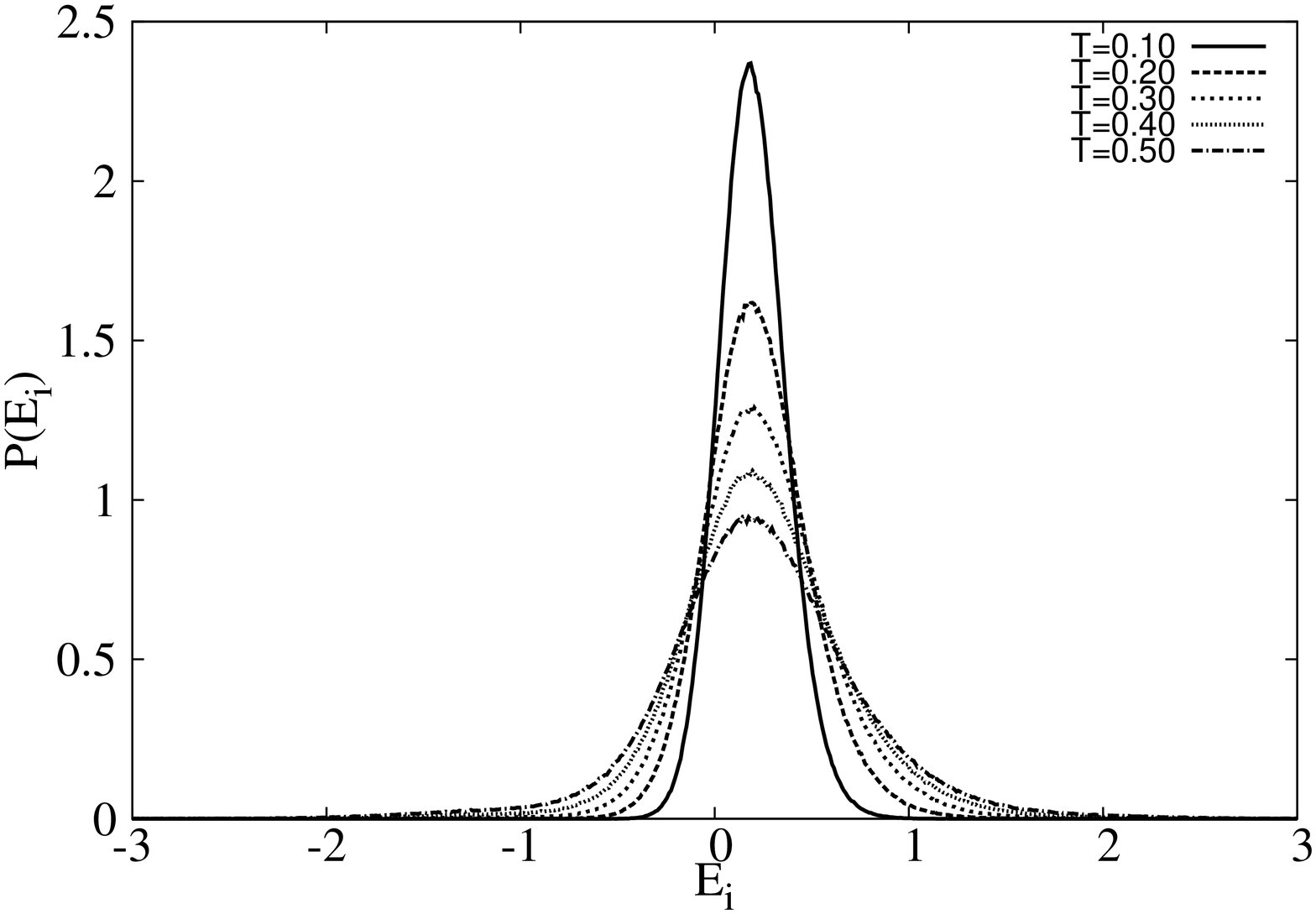}}
\hspace{0.01cm}
\caption{Plot of $P(E_i)$ at different values of $T=0$ for $\gamma=0.05$ (a), 
$\gamma=0.1$ (b), $\gamma=0.2$ (c) and $\gamma=0.5$ (d); $\tau = 8.0$, $F_0 = 0.2$.}
\label{fig:edge}
\end{figure}

This shows that the two dynamical states 
accessible to the particle for lower damping can be associated with
the regular trajectories in the deterministic regime. Incidently, for $\gamma = 0.12$, 
for the range of $0.18\le F_0\le 0.28$, two regular trajectories exists and it corresponds
exactly to the range of $F_0$ in which SR behaviour is observed in Ref. \cite{Reen1}.
The presence (or absence) of different number of dynamical states of trajectories in the
deterministic limit or for lower temperatures obviously depends on the value of friction
coefficent $\gamma$ and the amplitude of external drive $F_0$ under consideration.

\subsection{Input energy distributions}

\par Fig. 6 shows the input energy distributions for the different $\gamma$ values 
considered in Fig. 1  corresponding to different temperatures. The nature of the input 
energy distributions have earlier been succesfully used to understand the nature of the 
SR behaviour \cite{Saikia, Saikia1}.

The bimodal nature of the distributions at low temperatures for $\gamma = 0.05, 0.1$ and 
$0.12$ clearly shows the existence of the high energy and low energy dynamical states. 
As temperature rises the input energy distributions becomes asymmetrical, with highest 
asymmetry at temperatures corresponding to the SR peak. 
For very high temperatures, the input energy distributions again becomes symmetrical
about the mean value. However, for values of $\gamma$ for which SR behavior is absent, 
$\gamma = 1.5$, the input energy distributions are almost symmetrical for all values of 
temperature.

\subsection{Variation of $\bar\phi$ vis a vis $\langle \overline{E_{i}} \rangle$ with $\gamma$}

 \begin{figure}[h]
  \centering
  \subfigure[]{\includegraphics[width=5.5cm,height=6.7cm,angle=-90]{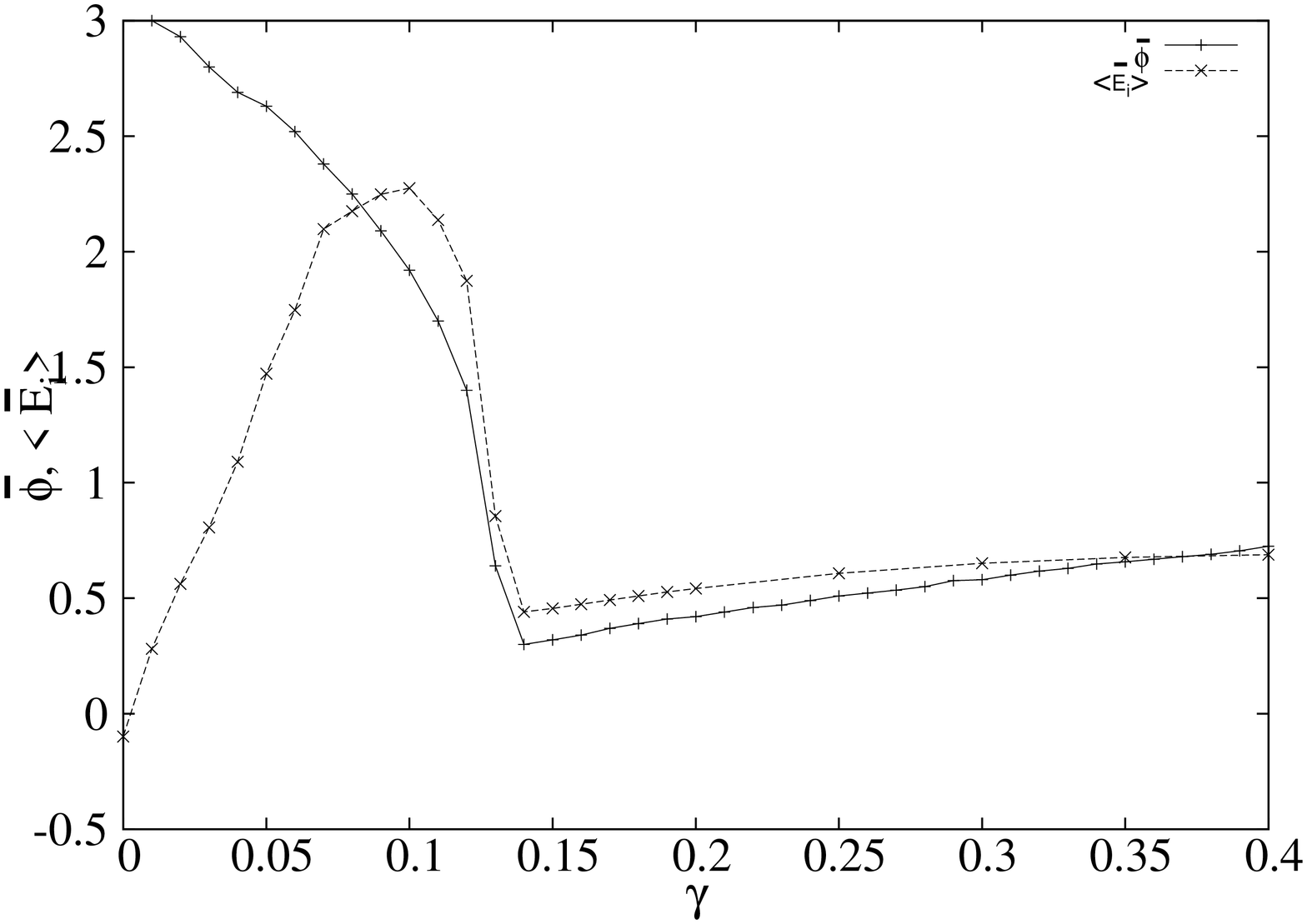}}
\hspace{0.01cm}
\subfigure[]{\label{fig:edge-c}\includegraphics[width=5.5cm,height=6.7cm,angle=-90]{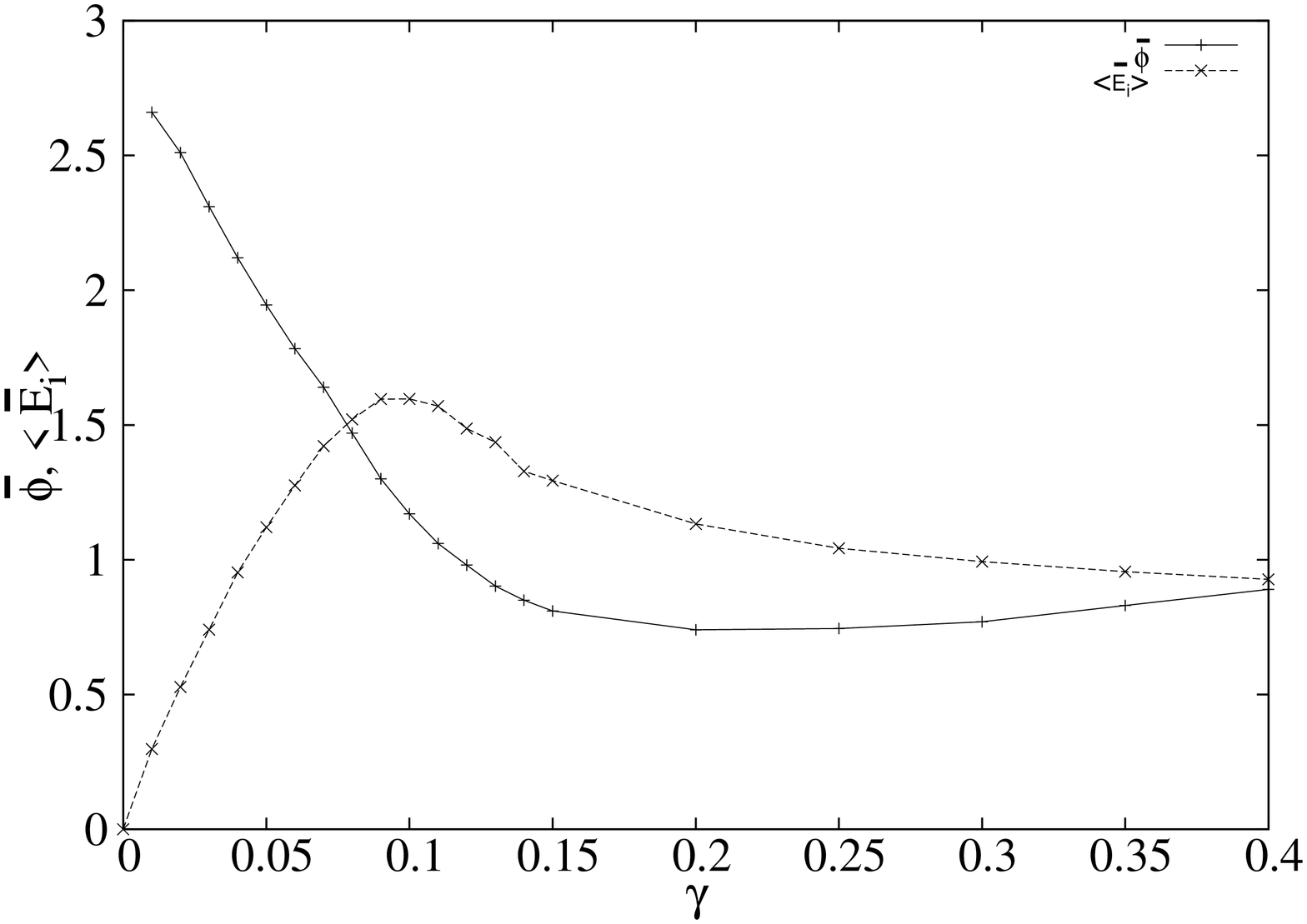}}
\caption{Plot of $\overline \phi$ and $\langle \overline{E_{i}} \rangle$ (increased by a 
factor of 4.0)  versus $\gamma$, for $T=0.0$ (a) and $T=0.15$ (b); $F_0 = 0.2$ and $\tau =8$.}
\label{fig:edge}
\end{figure}

In this section we study the variation of the phase difference $\bar{\phi}$ between the 
external drive $F(t)$ and the average particle response 
$\bar {x(t)}=\bar x_0 cos(\omega t +\bar \phi)$, and $\langle \overline{E_{i}} \rangle$
 with $\gamma$. It is seen that in the 
deterministic limit (Fig. 7a) and also for a finite temperature (Fig. 7b), $\bar \phi$ 
and $\langle \overline{E_{i}} \rangle$ shows a non monotonic behaviour with the change of
$\gamma$. With the increase of $\gamma$ from lower value, $\langle \overline{E_{i}} \rangle$ 
shows a peaking behaviour with $\gamma$, while $\bar{\phi}$ decreases showing an inflection 
point around the value of $\gamma$ where $\langle \overline{E_{i}} \rangle$ peaks. 
In Ref. \cite{Reen2, Saikia, Reen1, Gamma1} $\bar{\phi}$ was shown to have an inflection 
point at the peak of SR. We
observe that even with the variation of $\gamma$, $\langle \overline{E_{i}} \rangle$, 
exhibits a peaking with an associated inflection point in the phase lag.
\par In the deterministic case (Fig. 7a), as $\gamma$ changes, the average input energy
after attaining a peak again goes to a minimum. The value of $\gamma$ at the minima
corresponds to the level of damping at which the two dynamical states of trajectories
ceases to exist (as discussed above in the explanation of Fig. 2.

\section {Discussion and conclusion}
An underdamped periodic potential system exhibits SR when driven with a high frequency 
drive close to the natural frequency at the bottom of the potential wells. SR with average input 
energy per period $\langle \overline{E_{i}} \rangle$ as a quantifier, is found to occur 
only in the low damping regime. The transitions between the two definite states of the particle 
trajectories, characterised by their phase difference with the external drive, offers a 
possible explanation for the occurrence of SR \cite{Reen2, Saikia, Reen1, Liu}. 
The existence of these two states are found to be dependent on the choice of the friction 
coefficient $\gamma$ and the amplitude of drive $F_0$.
For $F_0=0.2$ considered here, in the low damping regime, $0 < \gamma < 0.14$, 
the two dynamical states exists at $T=0$ or at lower temperatures. However for 
$0.14 < \gamma < 1.5$, though, only a single dynamical state of trajectory exists near about
the deterministic limit, SR behaviour is observed. This is because, with the rise of 
temperature, the particle trajectories alternates between a high amplitude and low
amplitude regime. Synchronised transitions between these two regimes at an optimum temperature 
gives rise to the peak in $\langle \overline{E_{i}} \rangle$. The percentage of time that 
the particle spends in the high amplitude and low amplitude regime and its relation to SR 
needs to be further explored. As with temperature, the average input energy also shows
a peaking behaviour with the friction coefficient $\Gamma$. The phase difference $\bar\phi$ 
between $F(t)$ and $\bar x(t)$ has an inflection point exactly at the $\gamma$ value 
corresponding to the peak of $\langle \overline{E_{i}} \rangle$ vs $\gamma$ curve. So 
the criteria of the presence of an inflection in $\bar\phi$ is not exclusive only for 
SR peaks \cite{Reen2, Saikia, Reen1, Gamma1}.


\begin{thebibliography}{150}


\bibitem{Ben} R. Benzi, A. Sutera, and A. Vulpiani J. Phys. A 14, L453 
(1981).
\bibitem{Gam} L. Gammaitoni, P. H\"{a}nggi, P. Jung, and F. Marchesoni,
Rev. Mod. Phys. 70, 223 (1998).
\bibitem{Well} T. Wellens, V. Shatokhin, and A. Buchleitner, Rep. Prog. Phys.
67, 45 (2004).
\bibitem{Moss}K. Wiesenfeld, and F. Moss, Nature 373, 33 (1995).
\bibitem{Maddox} J. Maddox, Nature (London) 369, 271 (1994).
\bibitem{Hanggi} P. Hanggi, ChemPhysChem, 3, 285 (2002).
\bibitem{Ghosh} P.K. Ghosh, B.C. Bag, D.S. Ray, Phys. Rev. E 75, 032101 
(2007).
\bibitem{Douglass} J.K. Douglass, L. Wilkens, E. Pantazelou, and F. Moss,
Nature 365, 337 (1993).
\bibitem{Collins} J.J. Collins, T.T. Imhoff, and P. Grigg, J. Neurophysiology
76, 642 (1996).
\bibitem{Gluckman} B.J. Gluckman, T.I. Netoff, E.J. Neel, W.L. Ditto, M.L. 
Spano, nad S.J. Schiff, Phys. Rev. Lett. 77, 4098 (1996).
\bibitem{Simonotto} E. Simonotto, M. Riani, C. Seife, M. Roberts, J. Twitty,
and F. Moss, Phys. Rev. Lett. 78, 1186 (1997).
\bibitem{Shuhei} S. Ikemoto, F.D. Libera, K. Hosodo, Phys. Rev. E, 85, 021905 (2012).
\bibitem{Fuave} S. Fauve, and F. Heslot, Phys. Lett. A 97, 5 (1983).
\bibitem{Mantegna} R.N. Mantegna, and B. Spagnolo, Phys. Rev. E 49, R1792 
(1994).
\bibitem{Murali} K. Murali, S. Sinha, W.L. Ditto, A.R. Bulsara, Phys. Rev. 
Lett. 102, 104101 (2009).
\bibitem{Roy} B. McNamara, K. Wiesenfeld, and R. Roy, Phys. Rev. Lett. 60,
2626 (1988).
\bibitem{Mohanty} R.L. Badzey, and P. Mohanty, Nature 437, 995 (2005).

\bibitem{Stocks} N.G. Stocks, P.V.E. McClintock, and S.M. Soskin, Europhys.
Lett. 21, 395 (1993);
N.G. Stocks, N.D. Stein, and P.V.E. McClintock, J. Phys. A: Math. Gen. 26, 
L385 (1993).
\bibitem{Dykman} M.I. Dykman, D.G. Luchinsky, R. Mannella, P.V.E. McClintock,
 N.D. Stein, and N.G. Stocks J. Stat. Phys. 70, 479 (1993).

\bibitem{Arathi} S. Arathi and S. Rajasekar, Physica Scripta, Vol. 84, No. 6,
 065011 (2011).
\bibitem{Wies} K. Wiesenfield, D. Pierson, E. Pantazelou, C. Dannes and F. Moss,
Phys. Rev. Lett. 72, 2125 (1994).
\bibitem{Kauf} I.Kh. Kaufman, D.G. Luchinsky, P.V.E. McClintock, S.M. Soskin,
and N.D. Stein, Phys. Lett. A 220, 219 (1996).
\bibitem{Lin} N. Lin-Ru, G.Yu-Lan, M. Dong-Cheng, Chin. Phys. Lett. Vol. 26, No. 10
 (2009) 100505.
\bibitem{Dan} D. Dan, M. C. Mahato and A. M. Jayannavar, Phys. Lett. A 258,
217(1999), Phys. Rev. E 60, 6421 (1999).

\bibitem{Fron} L. Fronzoni and R. Mannela, J. Stat. Phys. 70, 501 (1993).
\bibitem{Hu}G. Hu, Phys. Lett. A, Vol. 174, Issue 3 (1993), 247.


\bibitem{Cass} J. M. Cassado, J. J. Mejias and M. Morillo, Phys. Lett. A 365, 366
 (1995).
\bibitem{Gitt} M. Gitterman, I. B. Khalfin and B. Ya. Shapiro, Phys. Lett. A, 339,
340 (1994).
\bibitem{Bao} J. D. Bao, Phys. Rev. E 62, 4606 (2000); Phys. Lett. A, Vol. 265, Issue 4,
244 (2000).
\bibitem{Marc} F. Marchesoni, Phys. Lett. A 231, 61 (1997).
\bibitem{Kall} J. Kallunki, M. Dube and T. Ala-Nissila, J. Phys. Cond. Mat. 11,
 9841 (1999).

\bibitem{Kim} Y.W. Kim, W. Sung, Phys. Rev. E 57 (1998) R6237.
\bibitem{Reen2}W.L. Reenbohn, S.S. Pohlong, M.C. Mangal, Phys. Rev. E 85 (2012) 031144. 
\bibitem{Schia} M. Schiavoni, F.-R. Carminati, L. Sanchez-Palencia, F. Renzoni and
 G. Grynberg, Eur. Phys. Lett., Vol. 59, No. 4, 493(2002).

\bibitem{Zhang} X. Zhang, J. Bao, Surface Science, Vol. 540, Issue 1, (2003) 145.
\bibitem{Saikia}S. Saikia, A.M. Jayannavar, M.C. Mahato, Phys. Rev. E 83 (2011) 061121.
\bibitem{Pareek} M. C. Mahato, T. P. Pareek and A. M. Jayannavar, Int. J. of Mod. Phys. B10, 3857(1996).
\bibitem{DanN} D. Dan, A. M. Jayannavar and M. C. Mahato, Int. J. of Mod. Phys. B14, 1585(2000).
\bibitem{SaikiaN} S. Saikia and M. C. Mahato, J. Phys.: Condens. Matter 21 (2009) 175409.

\bibitem{Reen1}W. L. Reenbohn, M. C. Mangal, Phys. Rev. E 88, 032143 (2013). 

\bibitem{Liu} K. Liu, Y. Jin, Physica A 392 (2013) 5283–5288.


\bibitem{Risk}H. Risken, {\it{The Fokker-Planck Equation}}(Springer, Berlin, 1989).
\bibitem{Fulde}P. Fulde, L. Pieternero, W. R. Schneider and S. Strassler, Phys. Rev.
Lett. 35, 1776 (1975);A. Asaklil, Y. Boughaleb, M. Mazroui, M. Chhib and L. El Arroum
, Solid State Ionics 159, 331 (2003).
\bibitem{Laca} A. M. Lacasta, J. M. Sancho, A. H. Romero, I. M. Sokolov and K. Lindenberg,
Phys. Rev. E, 70, 051104 (2004); A. P. Graha, F. Hoffmann, J. P. Toennies, L. Y. Chen
and S. C. Ying, Phys. Rev. B 56, 10567 (1997); D. C. Senft and E. Ehrlich, Phys. Rev.
Lett. 74, 294 (1995).
\bibitem{Falco}C. M. Falco, Am. J. Phys. 44, 733 (1976); A. Barone and G. Patterno,
Physics and Applications of the Josephson Effect (Wiley, Newyork, 1982).
\bibitem{Seki} K. Sekimoto, J. Phys. Soc. Jpn. 66, 1234 (1997).
\bibitem{Iwai}T. Iwai, Physica a 300, 300 (2001); T. Iwai, J. Phys. Soc. Jpn. 70, 353
(2001).
\bibitem{Gamma1} L. Gammaitoni, F. Marchesoni, M. Martinelli, L. Pardi and S. Santucci,
Phys. Lett. A 158, 449 (1991).
\bibitem{Dykman1} M. I. Dykman, R. Mannella, P. V. E. McClintock and N. G. Stocks,
Phys. Rev. Lett. 68, 2985 (1992); L. Gammaitoni and F. Marchesoni, Phys. Rev. Lett.
70, 873 (1993), M. I. Dykman, R. Manella, P. V. E. McClintock and N. G. Stocks, Phys.
Rev. Lett. 70, 874 (1993).

\bibitem{Dan1} D. Dan, A. M. Jayannavar, Physica A 345, 404 (2005).
\bibitem{Saikia1} S. Saikia, R. Roy, A. M. Jayannavar, Phys. Lett. A 369, 367 (2001)

\bibitem{Saikia2} S. Saikia, and M.C. Mahato, Physica A 389, 4052 (2010), and
references therein.
\bibitem{Mateos} J. L. Mateos, Phys. Rev. Lett. 84, 258 (2000).

\bibitem{Nume} R.~Mannella, A Gentle Introduction to the Integration
of Stochastic Differential Equations. In : {\it Stochastic Processes in 
Physics, Chemistry, and Biology}. Edited by J. A.~Freund and T. P\"{o}schel, 
Lecture Notes in Physics, vol. 557, 353. Springer, Berlin, 2000.

\bibitem{Mateos1}J. L. Mateos, Phys. Rev. Lett., 84, 258 (2000).





\end{thebibliography}
\end{document}